\documentclass[%
 reprint,
superscriptaddress,
 amsmath,amssymb,
 aps,
 twocolumn,
pra,
nofootinbib
]{revtex4-1}
\usepackage{amsmath,gensymb,textcomp,bm,dcolumn,eurosym,array,tabu,multirow,nicefrac,color,subfigure,graphicx,upgreek}
\usepackage[colorlinks, linkcolor=blue, citecolor=blue, urlcolor=blue, breaklinks=true]{hyperref}

\usepackage{times}

\usepackage{mathpazo,times} % Use a nicer font than the default

\newcommand{\ket}[1]{\lvert #1 \rangle}

\usepackage{graphicx}
\usepackage[table,xcdraw]{xcolor}
\usepackage{dcolumn}
\usepackage{units}
\usepackage{float}

\usepackage{fourier}
\DeclareRobustCommand{\bigO}{%
  \text{\usefont{OMS}{cmsy}{m}{n}O}%
}

\usepackage{color}
\usepackage[T1]{fontenc}
\usepackage[capitalize]{cleveref}
\renewcommand{\figurename}{\textbf{Fig.}}
\begin{document}

% Include your paper's title here

\title{A trusted-node-free eight-user metropolitan quantum communication network
}

% Include the date command, but leave its argument blank.

%\date{}
\author{Siddarth Koduru Joshi*}
\email{SK.Joshi@Bristol.ac.uk}
\affiliation{Quantum Engineering Technology Labs, H. H. Wills Physics Laboratory \& Department of Electrical and Electronic Engineering, University of Bristol}
%\thanks{*To whom correspondence should be addressed; E-mail:  SK.Joshi@Bristol.ac.uk.}
\author{Djeylan Aktas}
\affiliation{Quantum Engineering Technology Labs, H. H. Wills Physics Laboratory \& Department of Electrical and Electronic Engineering, University of Bristol}

\author{S\"oren Wengerowsky}
\affiliation{Institute for Quantum Optics and Quantum Information - Vienna (IQOQI) \& Vienna Center for Quantum Science and Technology (VCQ), Vienna, Austria}
\author{Martin Lon\v{c}ari\'{c}}
\affiliation{Photonics and Quantum Optics Research Unit, Center of Excellence for Advanced Materials and Sensing Devices, Ru\dj{}er Bo\v{s}kovi\'{c} Institute, Zagreb, Croatia}
\author{Sebastian Philipp Neumann}
\affiliation{Institute for Quantum Optics and Quantum Information - Vienna (IQOQI) \& Vienna Center for Quantum Science and Technology (VCQ), Vienna, Austria}
\author{Bo Liu}
\affiliation{College of Advanced Interdisciplinary Studies, NUDT, Changsha, 410073, China}
\author{Thomas Scheidl}
\affiliation{Institute for Quantum Optics and Quantum Information - Vienna (IQOQI) \& Vienna Center for Quantum Science and Technology (VCQ), Vienna, Austria}

\author{{Guillermo Curr\'as Lorenzo}}
\affiliation{School of Electronic and Electrical Engineering,University of Leeds, Leeds LS2 9JT, United Kingdom}

\author{\v{Z}eljko Samec}
\affiliation{Photonics and Quantum Optics Research Unit, Center of Excellence for Advanced Materials and Sensing Devices, Ru\dj{}er Bo\v{s}kovi\'{c} Institute, Zagreb, Croatia}
\author{Laurent Kling}
\affiliation{Institute for Quantum Optics and Quantum Information - Vienna (IQOQI) \& Vienna Center for Quantum Science and Technology (VCQ), Vienna, Austria}
\author{Alex Qiu}
\affiliation{Quantum Engineering Technology Labs \& Quantum Engineering Centre for Doctoral Training, Centre for Nanoscience and Quantum Information, University of Bristol, Bristol, United Kingdom}
\author{Mohsen Razavi}
\affiliation{School of Electronic and Electrical Engineering,University of Leeds, Leeds LS2 9JT, United Kingdom}
\author{Mario Stip\v{c}evi\'{c}}
\affiliation{Photonics and Quantum Optics Research Unit, Center of Excellence for Advanced Materials and Sensing Devices, Ru\dj{}er Bo\v{s}kovi\'{c} Institute, Zagreb, Croatia}
\author{John G. Rarity}
\affiliation{Quantum Engineering Technology Labs, H. H. Wills Physics Laboratory \& Department of Electrical and Electronic Engineering, University of Bristol}
\author{Rupert Ursin}
\affiliation{Institute for Quantum Optics and Quantum Information - Vienna (IQOQI) \& Vienna Center for Quantum Science and Technology (VCQ), Vienna, Austria}

%%%%%%%%%%%%%%%%% END OF PREAMBLE %%%%%%%%%%%%%%%%

%\begin{document} 

% Double-space the manuscript.

%\baselineskip24pt

% Make the title.

% Place your abstract within the special {sciabstract} environment.

\begin{abstract}
Abstract:  Quantum communication is rapidly gaining popularity due to its high security and technological maturity. However, most implementations are limited to just two communicating parties (users). Quantum communication networks aim to connect a multitude of users. Here we present a fully connected quantum communication network on a city wide scale without active switching or trusted nodes. We demonstrate simultaneous and secure connections between all 28 pairings of 8 users. Our novel network topology is easily scalable to many users, allows traffic management features and minimizes the infrastructure as well as the user hardware needed.
\end{abstract}

\maketitle

% In setting up this template for *Science* papers, we've used both
% the \section* command and the \paragraph* command for topical
% divisions.  Which you use will of course depend on the type of paper
% you're writing.  Review Articles tend to have displayed headings, for
% which \section* is more appropriate; Research Articles, when they have
% formal topical divisions at all, tend to signal them with bold text
% that runs into the paragraph, for which \paragraph* is the right
% choice.  Either way, use the asterisk (*) modifier, as shown, to
% suppress numbering.

\section*{Introduction}

Quantum communication networks present a revolutionary step in the field of quantum
communication~\cite{scarani2009security,pirandola2019advances}. Despite real world demonstrations of Quantum Key Distribution (QKD)~\cite{wengerowsky2018field,hipp2016demonstration,wang2015experimental,patel2012coexistence,qiu2014chinalink,wengerowsky2020passively}, the difficulty of scaling the standard two-user QKD protocols to many users has prevented the large scale adoption of quantum communication. Thus far, quantum networks relied upon one or more problematic features: trusted nodes~\cite{Peev2009,Sasaki2011,Stucki2011,Xu2009,FieldQKD_Wang14} that are a potential security risk, active switching ~\cite{Toliver2003,Chen:10-chinese-access-network,Elliott2005,Chang2016} which restricts both functionality and connectivity and most recently wavelength multiplexing~\cite{Soeren2018} with limited scalability.
The ultimate goal of quantum communication research is to enable widespread connectivity, much like the current internet, with security based on the laws of physics rather than computational complexity.
 To achieve this, a quantum network must be scalable, allow users with dissimilar hardware, be compatible with traffic management techniques, must not limit permitted network topologies and as far as possible avoid potential security risks like trusted nodes. 
 
 So far, all demonstrated QKD networks fall in three broad categories.
First, trusted node  networks~\cite{Peev2009,Sasaki2011,Stucki2011,Xu2009} where some or all nodes in a network are assumed to be safe {from eavesdropping}. In most practical networks, it is rare to be able to trust every connected node. Furthermore, such networks tend to utilize multiple copies of both the sender and receiver hardware at each node, thereby increasing the cost prohibitively.
Second, actively switched or ``access networks'' where only certain pairs of users are allowed to exchange a key at a time~\cite{Herbauts13}. Similarly point to multipoint networks are useful in niche applications and have been shown using passive beamsplitters~\cite{Townsend1997,Choi2011,Frohlich2013}, active switches~\cite{Toliver2003,Chen:10-chinese-access-network,Elliott2005,Chang2016}, and frequency multiplexing~\cite{Aktas2016,Chang2016,Zhu15,lim2008}.
Lastly, fully connected quantum networks which can be based on high-dimensional/multi-partite entanglement to share entanglement resources between several users~\cite{Torma1999,pivoluska2017layered}. However, the extreme complexity of changing the dimensionality of the state produced by the source makes this approach unscalable. Fortunately, fully connected networks (i.e. where every user is connected to every other user directly)  can be achieved using multiplexing and bipartite entanglement~\cite{Soeren2018}. Nevertheless, the scheme in Ref~\cite{Soeren2018} requires 
%$n(n-1)$ 
$\bigO(n^2)$
wavelength channels for $n$ users, which prevents the technology from being scaled to more than a few users.

Here we present a city-wide quantum communication network, with 8 users, that forms a  fully connected graph/network (where each user exchanges a secure key with every other user simultaneously) while requiring only 8 wavelength channel pairs, minimal user hardware (i.e., 2 detectors and a polarization analysis module), and no trusted nodes.
The quadratic improvement in resources (the number of channels) used is due to passively multiplexing using both wavelength filters and beamsplitters (BS). 
%In addition we could facilitate traffic management features by choosing, with an active component, how wavelength pairs are distributed. 
Further, to the best of our knowledge, we have demonstrated the largest quantum network without trusted nodes to date. 
Similar to Ref~\cite{Soeren2018}, just one  source of polarization entangled photon pairs is shared passively between all users and requires neither trust in the service provider nor adaptations to add or remove users.
%We go significantly beyond previous works as 
We performed a full QKD experiment on a city wide scale in deployed fibers with a mixture of Superconducting Nanowire Single Photon Detectors (SNSPDs) and a Single Photon Avalanche Diode (SPAD). Further, we demonstrate a new topology with 
$\bigO(n)$
%$\mathcal{O}(n)$ 
scaling of all resources consumed using only 16 wavelength and 2 BS channels to distribute 8 entangled states, among the 28 links,  between 8 users in a fully connected graph using only one fiber and polarization analysis module per user.

\section*{Results}

\begin{figure*}[]
    \centering
    \includegraphics[width=0.95\textwidth]{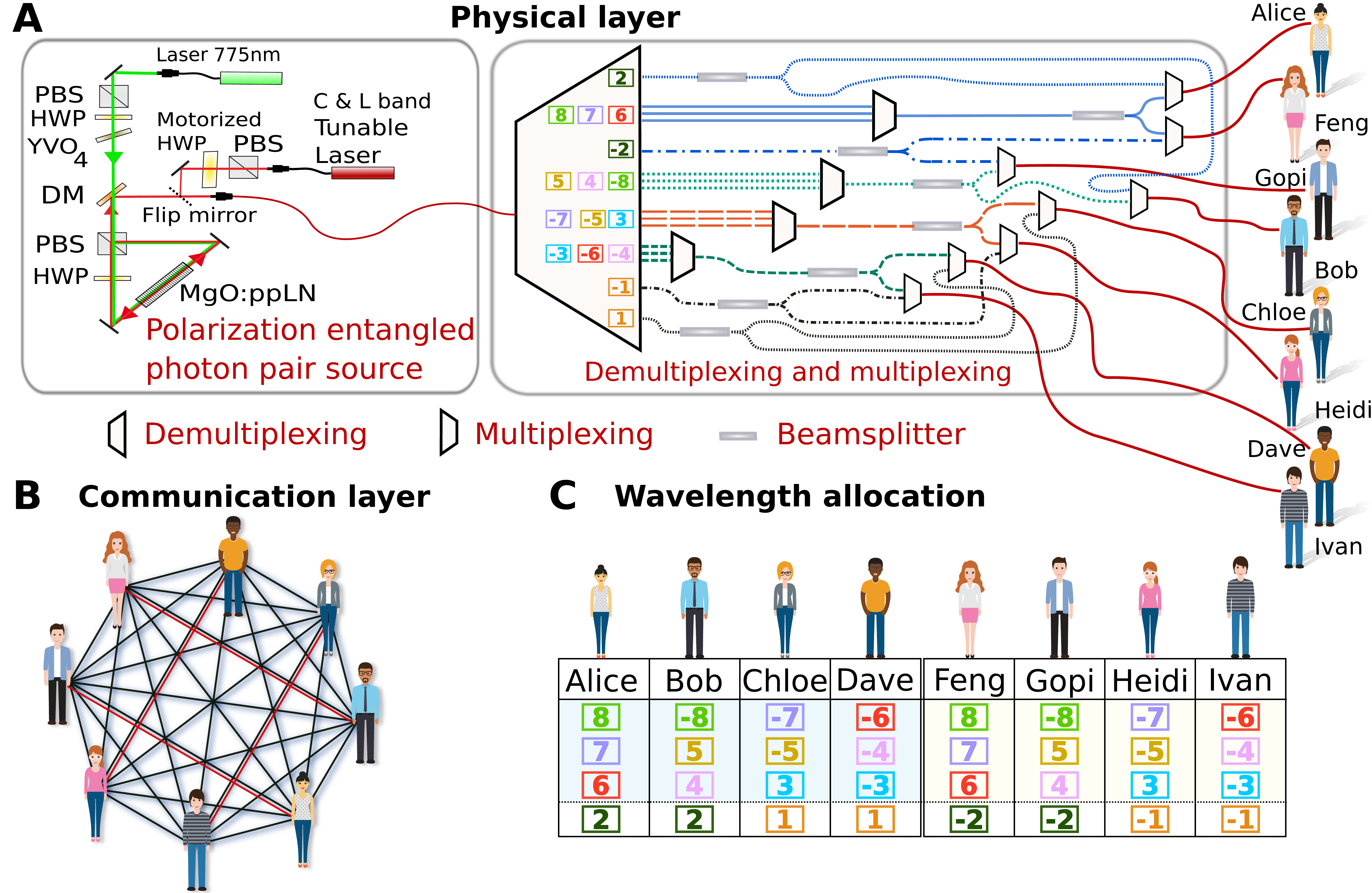}
    \caption{\small \textbf{The overall network architecture showing the physical layer, communication layer and the way wavelength channels are distributed}: The network consists of two layers. (\textbf{A}) \emph{The physical layer} contains the source of entanglement (blue) and the multiplexing unit (gray). These form the Quantum Network Service Provider (QNSP). Our topology uses just one deployed fiber (red) per user to interconnect all 8 individual users. (\textbf{B}) \emph{The communication layer} forms a fully connected graph without trusted nodes for: entanglement distribution, key exchange and secure communication (classical communication channels between users are not shown). Each line represents a link -- the sharing of a bipartite entangled state. Higher bandwidth links share a second entangled state shown as a red line. (\textbf{C}) \emph{Wavelength allocation}: Every user of the 8 node network receives 4 wavelength channels denoted by a number (which  corresponds to their ITU 100\,GHz DWDM grid channel number minus 34).    
I.e., ITU channel 34 (or 0 in the figures naming convention) corresponds to the channel approximately centered at the downconversion degeneracy wavelength. Thus, a pair of matching colors or numbers with equal absolute values and opposite sign denote wavelengths corresponding to an entangled photon pair. The regions shaded in blue and yellow are identical sub-nets and represent the multiplexing using beamsplitters. The last row below the dotted line shows the additional wavelength channel needed to fully interconnect the two sub-nets. Certain user pairs are connected by two entangled photon pairs (such as Alice and Gopi via \{8, -8\} and \{2, -2\}) and consequently enjoy an increased key rate. 
%By redistributing the channels shown in the above table, we could establish such ``premium'' links between any pair of users, thus allowing traffic management functionality.
    }
    \label{fig:layers}
\end{figure*}

\begin{figure*}[htpb]
\centering
  % \hspace{-10ex} 
   \includegraphics[width=0.95\textwidth]{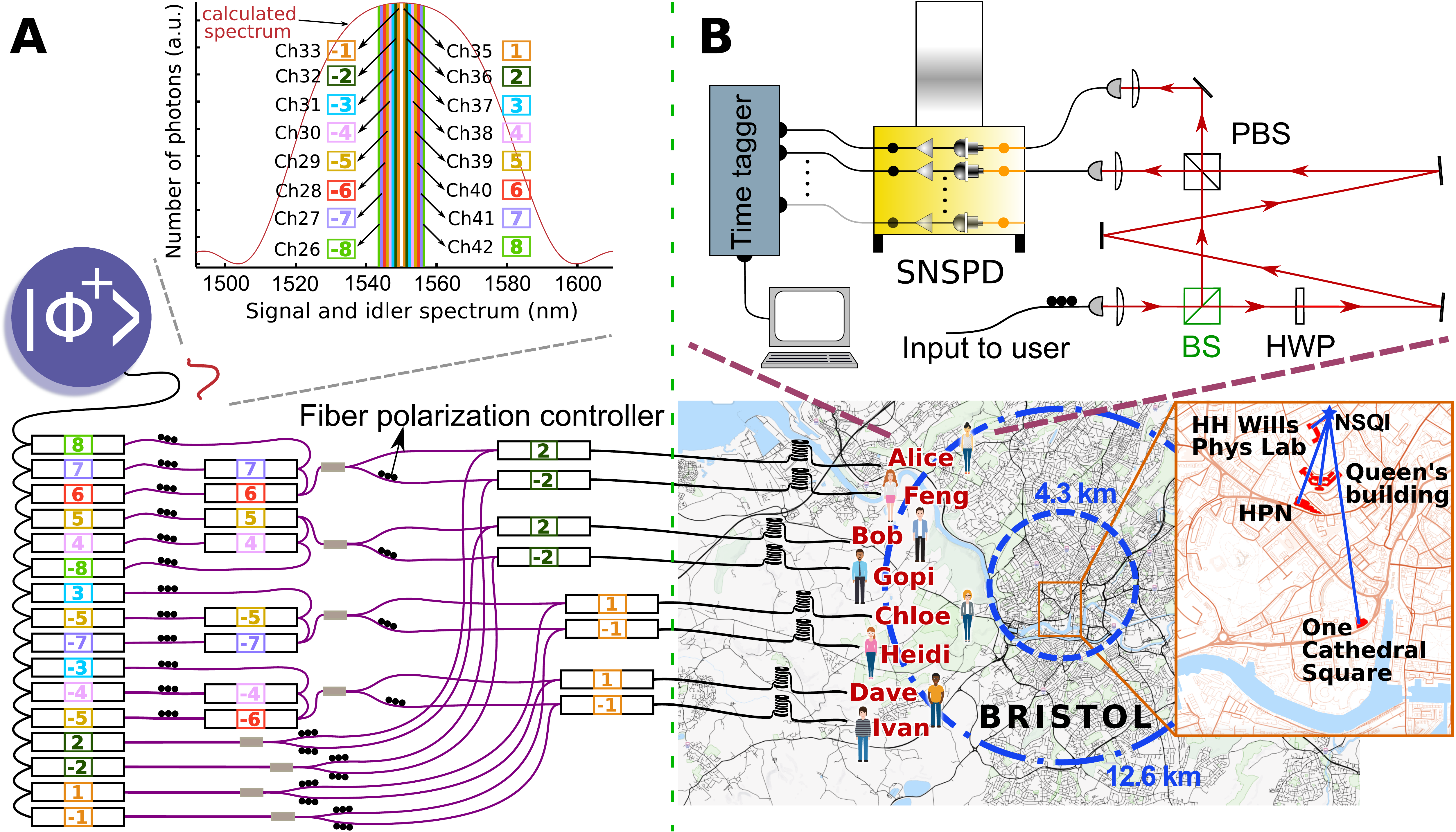}
    \caption{\textbf{The experimental setup showing the multiplexing and demultiplexing steps anlong with the user module and distribution of users across the city of Bristol}:  A source of bi-partite polarization entangled photon pairs with a broad band signal and idler spectrum (as shown in the inset (\textbf{A})) produces a $\ket{\phi^+}$  Bell state that is wavelength and BS multiplexed as shown. Wavelength multiplexing was performed using 100 GHz ITU DWDM channels represented as coloured numbers plus or minus the central channel 34.  BS multiplexing used 50:50 fiber beamsplitters (BS). Photons were sent to each user via loop backs from deployed fibers spread across the Bristol city centre (blue links in the above map) or several kilometers of fiber coil whose  effective coverage is shown by the blue dashed circles on the map. The measurement apparatus of each user is shown in the inset b). To the left of the dashed green line is the Quantum Network Service Provider (QNSP). Maps plotted using data from mappyplace.com and mapiful.com. (\textbf{A}) {The spectrum of the signal and idler photons} was calculated from data-sheet values and Sellmeier coefficients of  Ref~\cite{gayer2008sellmeier}. Energy conservation ensures that pairs of wavelengths when  at the same spectral distance from the central wavelength, are correlated. Such a pair of wavelength channels is indicated by the same color number with and without the minus sign. The ITU channels numbers along with their representative colored numbers are shown. (\textbf{B}) {The Polarization Analysis Module} (PAM) of each user consists of a beamsplitter to direct input photons along either the short or long optical path. The short path measures the polarization in the HV basis using a polarizing beamsplitter (PBS) and two superconducting nanowire single photon detectors (SNSPD). The long path includes an achromatic half-wave plate to rotate the measurement basis to DA and measures using the same PBS and SNSPDs. 
    %We note that the long and short paths are incident on the same detector but are not aligned to be an interferometer. 
    }
    \label{fig:source+wdms}
\end{figure*}

By using a combination of standard telecom Dense Wavelength Division Multiplexers (DWDM) with 100\,GHz channel spacing and BS multiplexing using in-fiber beamsplitters,
 we were able to distribute bi-partite entangled states between all users from just one source of polarization entangled photon pairs. The network architecture requires only 16 wavelength channels to fully interconnect 8 users as opposed to the 56 channels that would be necessary following our earlier scheme~\cite{Soeren2018}. 
 The network architecture is best understood when divided into different layers of abstraction as shown in Fig.~\ref{fig:layers}. The bottom ``physical layer'' represents the actual infrastructure that supports the network and comprises of a central Quantum Network Service Provider (QNSP) and the user hardware connected to the QNSP via distribution fibers.
 In the physical layer, the network topology requires only one fiber between each user and the service provider, while in the logical/connection layer the topology naturally forms a fully connected graph between all 28 unique pairs formed by 8 users (see Fig.~\ref{fig:layers}). Every user is equipped with a polarization analysis module that implements a passive basis choice using two single photon detectors each as shown in Fig.~\ref{fig:source+wdms}. Users can  demultiplex the incident photons such that each detector receives fewer wavelength channels to improve their signal-to-noise ratio and therefore the key rate. 
  We generate secure keys between all 28 links formed by pairs of users. Four of these links can be chosen to have \emph{premium} connections with increased key rates, which when combined with active switching can provide traffic management on the network. Lastly, we demonstrate that our network is capable of supporting a mixture of both SNSPD and SPAD based user platforms.
  
 The details of the QNSP are shown to the left of Fig.~\ref{fig:source+wdms}. It consists of both the source of polarization entangled photon pairs and the Multiplexing Unit (MU) comprising WDMs and beamsplitters. 
All multiplexing is performed in a single MU, co-located with the source in our implementation. To take advantage of existing fiber infrastructure, channels for many users can be sent along fewer fibers and multiple MUs, at various locations closer to clusters of users, can be used to create this quantum network.
The user hardware -- a Polarization Analysis Module (PAM) and two single photon detectors -- is shown in  Fig.~\ref{fig:source+wdms} inset b).
Photons incident on a PAM are directed by a beamsplitter along either the short path where they are measured in the horizontal/vertical polarization (HV) basis, or along the long path and through a half-wave plate such that they are measured in the diagonal/anti-diagonal polarization (DA) basis. 
The overall result is that the physical layer constitutes a relatively simple hub and spoke topology,  while in the logical layer, every pair of users always share an entangled photon pair.

We conceptually divide the 8 users of our network referred to as Alice (A), Bob (B), Chloe (C), Dave (D), Feng (F), Gopi (G), Heidi (H) and Ivan (I), into two sub-nets of 4 users (see Fig.~\ref{fig:layers}). A sub-net uses wavelength multiplexing to form a fully connected network among its members -- A, B, C, D. Beamsplitters are then used in each of the wavelength channels to duplicate the first sub-group creating a second set of 4 interconnected users -- F, G, H, I.
Thus entanglement is shared between every pair of users except AF, BG, CH and DI as the above splitting also gives rise to connections between the two sets. Two additional wavelength pairs are then distributed between these pairs of users to create the fully connected network of 8 users with only 16 wavelength channels. 
Each pair of users performs a standard BBM92~\cite{BBM92} protocol where the photons shared with all other users are treated as background noise. A narrow coincidence window, optimized in post processing and typically about 130\,ps,  ensures that this noise only contributes minimally to the Quantum Bit Error Rate (QBER).

%\section{\label{sec:Setup}Setup}

All multiplexing and demultiplexing in the experiment is performed with standard telecommunications equipment.
The experimental setup (shown in Fig.~\ref{fig:source+wdms}) uses a broadband source of polarization entangled photon pairs at telecommunications wavelengths similar to that described in Ref~\cite{Soeren2018}. Comparable sources have also been reported in Refs.~\cite{lim2008,Herbauts13,Aktas2016,Zhu:19,Chen:18} where a $\sim${775}\,nm pump downconverts in a Type 0 MgO:PPLN crystal to produce signal and idler photons with a Full Width at Half Maximum (FWHM) bandwidth of $\sim${60}\,nm centered at {1550.217 nm}  (roughly corresponding to the International Telecommunication Union's (ITU) 100\,GHz DWDM grid at channel 34 (Ch34) centered at 1550.12\,nm). Due to energy conservation during the down conversion process, only frequencies equidistant from half the pump frequency can support entangled pairs. Thus, wavelengths corresponding to the channel pairs \{Ch33, Ch35\}, \{Ch32, Ch36\}, \{Ch31, Ch37\}, and so on are entangled with each other. The wide signal and idler spectrum {was} demultiplexed into 8 wavelength pairs as above and each user is given a combination of wavelengths according to the table in Fig.~\ref{fig:layers}. Thus each user receives 4 {wavelength} multiplexed channels simultaneously via one single mode fiber and  {8} different entangled states are shared between the 28 different pairs of users.

%\subsection{\label{sec:WDM}Demultiplexing and Multiplexing}

%\section{Results}\label{sec:results}

\begin{figure}[htb]
    \centering
    \includegraphics[width=0.95\columnwidth]{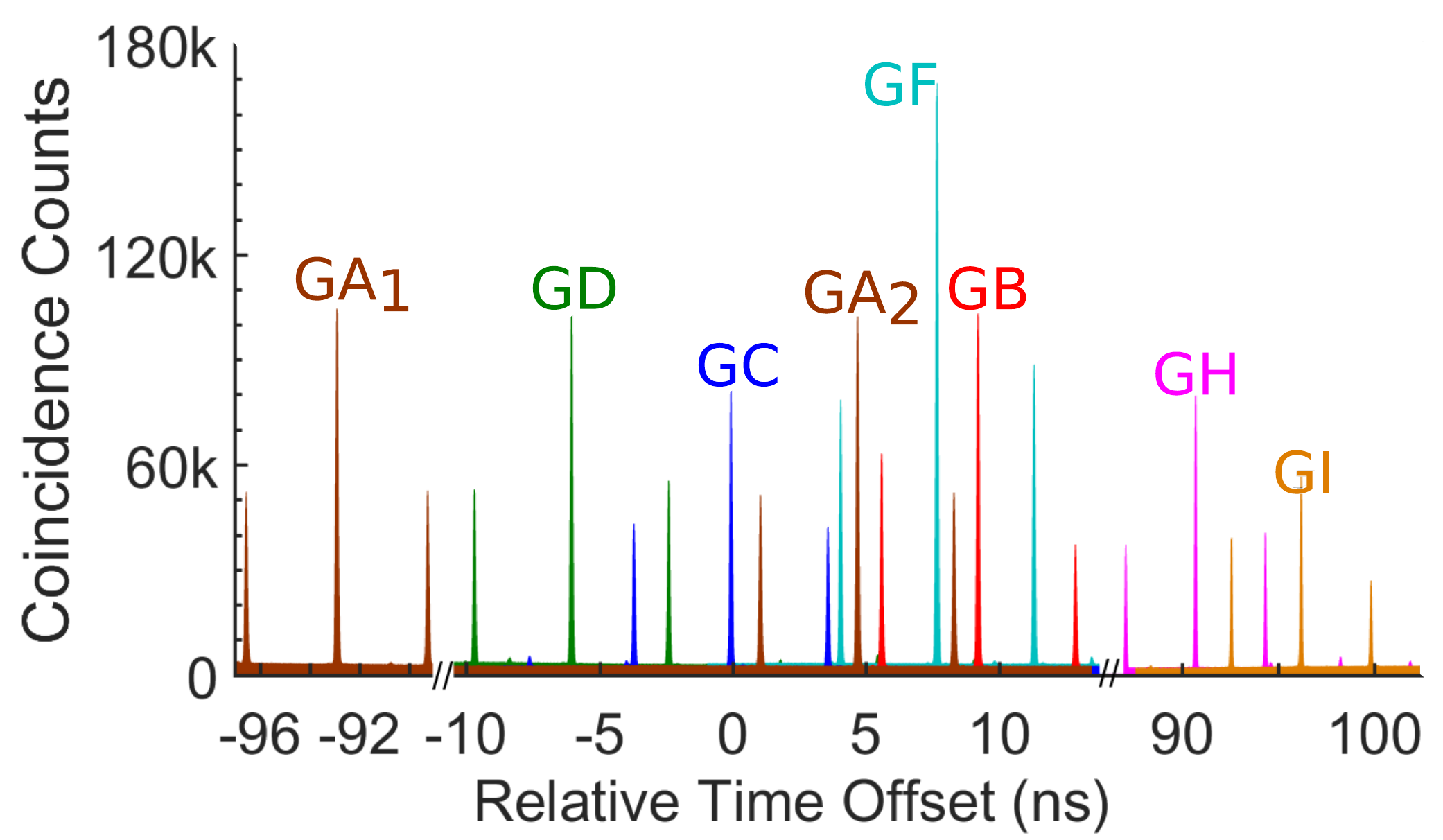}
    \caption{\textbf{Secure key rate over time for the lab experiment}. A secure key was generated every 10\,min while including finite key effects and a security parameter of {$10^{-5}$}. The length of each link is given in Table~\ref{tab:length}, while the average secure key rate for each link is tabulated Table~\ref{tab:long} for the metropolitan network and Table~\ref{tab:short} for the in laboratory demonstration.}
    \label{fig:stability} 
\end{figure}

The experiment was performed in two stages. In the first stage the QNSP, MU, the 8 users each connected to the  QNSP/MU with a single fiber $\sim$10\,m in length, and the 16 detectors were situated in a single laboratory in the Nano Science and Quantum Information (NSQI) building in Bristol.
To demonstrate the stability of our network we recorded data for {18.45}\,hours as shown in
Fig.~\ref{fig:stability}. 
To be able to account for finite key effects with a security parameter of {$10^{-5}$}, we computed the private key once every 10 minutes and the figure shows the average secure key generation rate per second in each 10\,min period for each of the 28 links (discussed further in the methods section). 
The total secure key obtained is shown in Table~\ref{tab:short}. 
Users A through H used superconducting nanowire detectors from Photon Spot while Ivan used a combination of one SNSPD and one InGaAs single photon avalanche photodiode (SPAD). We note that the use of heterogeneous detectors did not significantly impact the key generation rates.

\begin{table}[htb]
%\centering

    \begin{tabular}[b]{c|c|c|c|c|c|c|c|c}
  & Bob & Chloe & Dave & Feng & Gopi & Heidi & Ivan\\\hline
Alice   & 10.03 & 10.08
 & 9.58 & 13.37 & 16.53 & 9.06 & 6.81\\
Bob  &  & 9.14 & 8.58 & 17.32 & 6.24 & 7.44 & 5.95\\
Chloe  & &  & 14.25 & 12.64 & 7.01 & 8.67 & 14.63 \\
Dave  & & &  & 10.64 & 6.33 & 20.27 & 11.45
 \\
Feng  & & & &  & 9.01 & 10.80 & 4.44 \\
Gopi  & & & &  &  & 5.96 & 3.88 \\
Heidi  & & & & &  &  & 6.43 \\
%Ivan & & & & & & &  &  \\
\hline
\end{tabular}

\raggedright
\caption{\textbf{Total secure key (Mega bits) for the laboratory demonstration}  as measured continuously over 18.45 hours after accounting for all finite key size effects.  }
    \label{tab:short}
\end{table}

In the second stage the connection between the user and the MU was replaced for 6 users by long distance links. Furthermore, the SPAD was exchanged between Ivan and Gopi. Alice was connected via a 12.6\,km spool with a loss of 13.3\,dB, Chloe was connected to a loop-back from a laboratory in the first floor of the physics building of the University of Bristol with a total distance of 463\,m and 1.36\,dB loss, Dave used a 4.3\,km spool (15.7\,dB), Feng looped back from the basement of the Merchant Ventureres Building (MVB) through 1.625\,km of deployed fiber (and a loss of 2.04\,dB), Heidi utilized a loop-back connection from the ground floor of Queen's Building with a loss of 1.68\,dB and a total distance of 1.624\,km, Ivan was connected to another loop-back from the server room of One Cathedral Square in the city center totaling 3.10\,km (2.57\,dB). Bob and Gopi continued to be connected via short fibers. Thus, the 28 links varied in the effective separation of users from 16.6\,km to $\sim$10\,m. This shows the versatility of the network architecture as both a local area network and a city wide metropolitan area network.
Table \ref{tab:long} shows the secure key rate over these long distances in deployed fiber and in fiber spools.

\begin{table}[tb]
\centering
    \begin{tabular}[b]{c|c|c|c|c|c|c|c|c}
\hline
 & Bob & Chloe & Dave & Feng & Gopi & Heidi & Ivan\\\hline
Alice &  31143
 & 8926
 & 6087
 & 15590
 & 38075
 & 6637
 & 901
\\
Bob  &  & 23747
 & 24171
 & 83986
 & 41239
 & 9380
 & 9787
\\
Chloe  & &  & 16850
 & 14842
 & 17910
 & 9511
 & 12694
 \\
Dave  & & &  & 1516
 & 17004
 & 14230
 & 4356
 \\
Feng  & & & &  & 20121 & 10142 & 810 \\
Gopi  & & & & &  & 9954
 & 3759
 \\
Heidi  & & & & & &  & 1747
 \\
\hline
\end{tabular}

\raggedright
\caption{\textbf{Total secure key (bits) over long distance links for the city wide metropolitan network demonstration.} We connected 4 locations/users across the city of Bristol as shown in Fig.~\ref{fig:source+wdms} via deployed fiber in a loop back configuration. Two other users were sent signals through fiber spools and the remaining two were connected via short (10\,m) fibers. The distances of all 28 links are given in Table~\ref{tab:length}. Considering finite-size-effects, we measured for $\sim$27 minutes to obtain the final secure key shown. Here, we set  the failure probability of phase error estimation to $10^{-5}$. Fig.~\ref{fig:Keystability} shows the overall stability of the key for 7\,hours.}

%\raggedright
%    {}
    \label{tab:long}
\end{table}

The QBER, and hence the secure key rate, in our proof of principle experiment was limited by two main experimental imperfections: Firstly, a more careful fiber neutralization procedure using the manual Fiber Polarization Controllers would significantly improve the QBER.
Secondly, the alignment of the HWP and extinction ratio of the polarizing beamsplitter (PBS) used in each user's PAM could be further optimized (see supplementary material).  
%Did not include accidentals because they are not an imperfection.

\begin{figure*}
    \centering

    \includegraphics[width=0.95\textwidth]{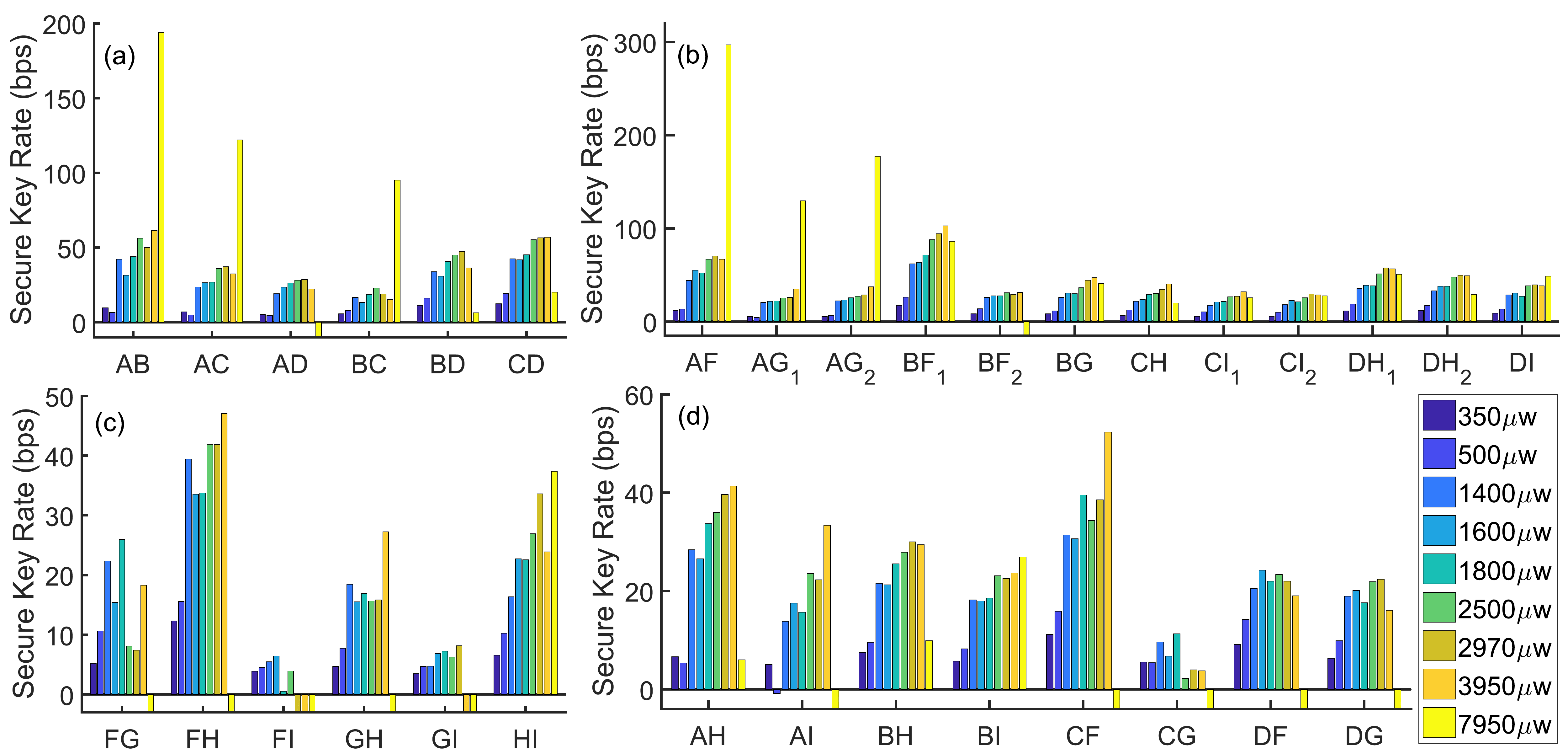}
    \caption{\textbf{Optimizing the average secure key rate by adjusting the pump power}: The amount of secure key obtained per second can be optimized by increasing the pump power at the source and hence the pair production rate.}
    \label{fig:pow}
\end{figure*}

Further optimization of the secure key rate is possible by adjusting the pump power in the source thus, changing the pair generation rate (see Fig.~\ref{fig:pow}). We cannot adjust the pair generation rate in each pair of wavelength channels separately. Thus the optimum pump power is strongly effected by the different types/alignment of user hardware (like detectors and PAMs) in the network.

\begin{figure*}[t!]
    %\centering
    \includegraphics[width=0.95\textwidth]{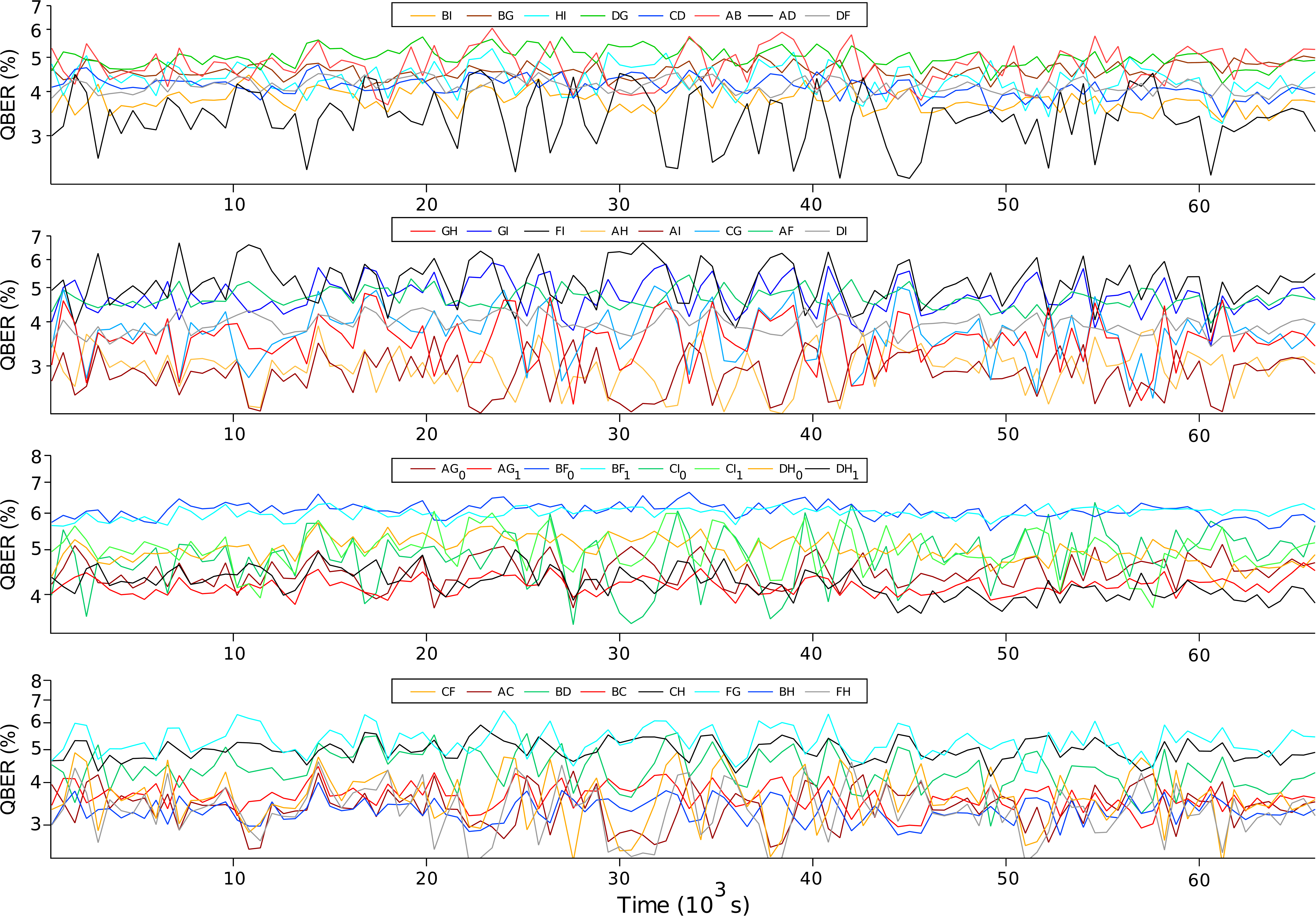}
    \caption{\textbf{Stability of the QBER over extended periods of time proves that polarisation encoding over fibers is a viable solution}. We tested the passive stability of our network over short links and a very long time of 18.45 hours. The QBER for each pair of users is shown here. In addition, those users with premium links (i.e. more than one set of correlated wavelengths) shared between them have two independent values of the QBER and are indicated using the subscripts 0,1. The secure key rate for this measurement is shown in Fig.~\ref{fig:stability}.
    \label{fig:qberstability}}
\end{figure*}

\begin{figure}
    %\centering
    \includegraphics[width=0.95\columnwidth]{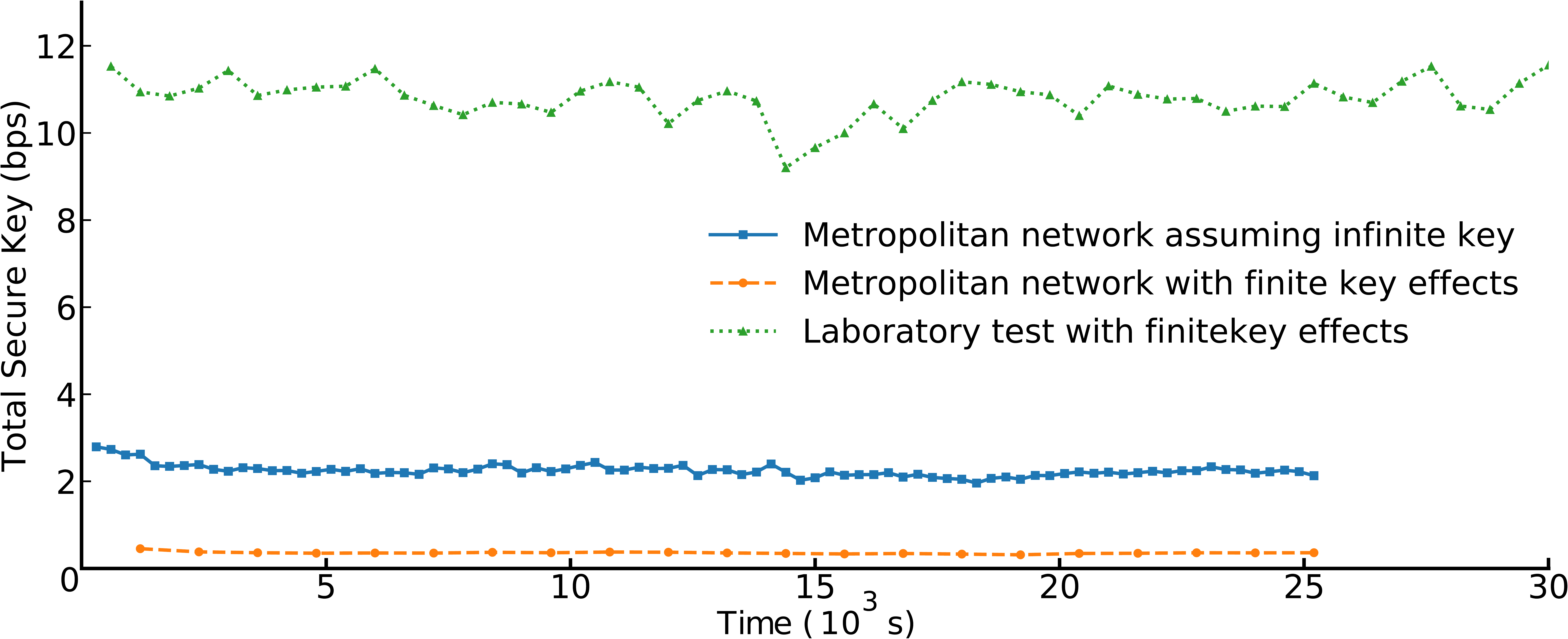}
    \caption{\textbf{Stability of the secure key rate over time}. To compare the test network in the laboratory with the real world deployed citywide fiber network, we summed up the  key rates from each pair of users. We note that despite the high losses and large distances involved (up to $\sim$17\,km) the network's key rate remains stable. The key rate in bits per second is shown while considering finite key effects for the Metropolitan quantum communication network (laboratory test) in blue solid line (green dotted line) using block sizes of 20\,min (10\,min). For comparison we also show the key rate of the city wide network assuming an infinite key length averaged over a block size of 5\,min.
    \label{fig:Keystability}}
    
\end{figure}

%An overview of the  
%operational parameters of the detectors used are found in the Methods section.
Nevertheless, the measured QBER proved to be stable in an 18.45\,hours laboratory test (see Fig.~\ref{fig:qberstability}) and resulted in a stable and positive overall secure key rate in a 7 hour metropolitan quantum communication network demonstration (see Fig.~\ref{fig:Keystability}).

\begin{figure*}
    \centering

    \includegraphics[width=0.95\textwidth]{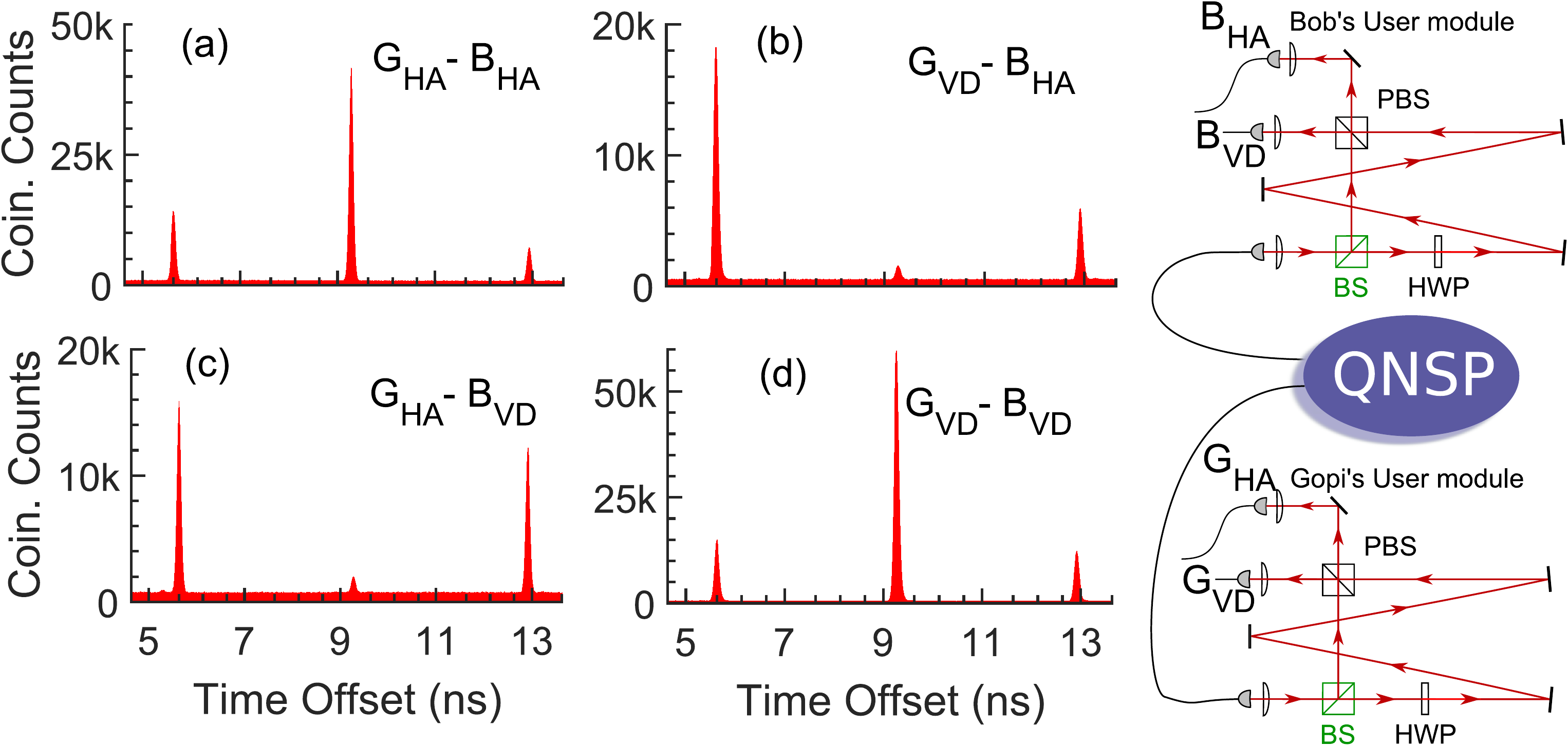}
    \caption{\textbf{Temporal cross correlation histograms between users Gopi and Bob illustrate how users generate quantum secure keys}: $g^{(2)}$ histograms between users Gopi (G) and Bob (B) are shown without obscuring the channel information. Each user's Polarization Analysis Module (PAM) detects photons in the Horizontal (H) or Anti-diagonal (A) basis on detector 1 and Vertical (V) or Diagonal (D) on detector 2.  D and A detection events are delayed with respect to H and V by $\sim$3.7\,ns. Four different histograms corresponding to each possible pair of detectors between G and B are shown. From this data we can directly measure the QBER by comparing the desired middle peaks (upper left \& lower right) with the undesired ones (upper right \& lower left). The data shown was integrated over one hour. 
    To the right is a simplified schematic showing two users connected to the Quantum Network Service Provider (QNSP) with the relevant detectors labeled. The peak separation of 3.7\,ns is primarily due to the optical path length difference is each user module. \label{fig:Histgb}}
\end{figure*}

\begin{figure*}
    \centering

    \includegraphics[width=0.95\textwidth]{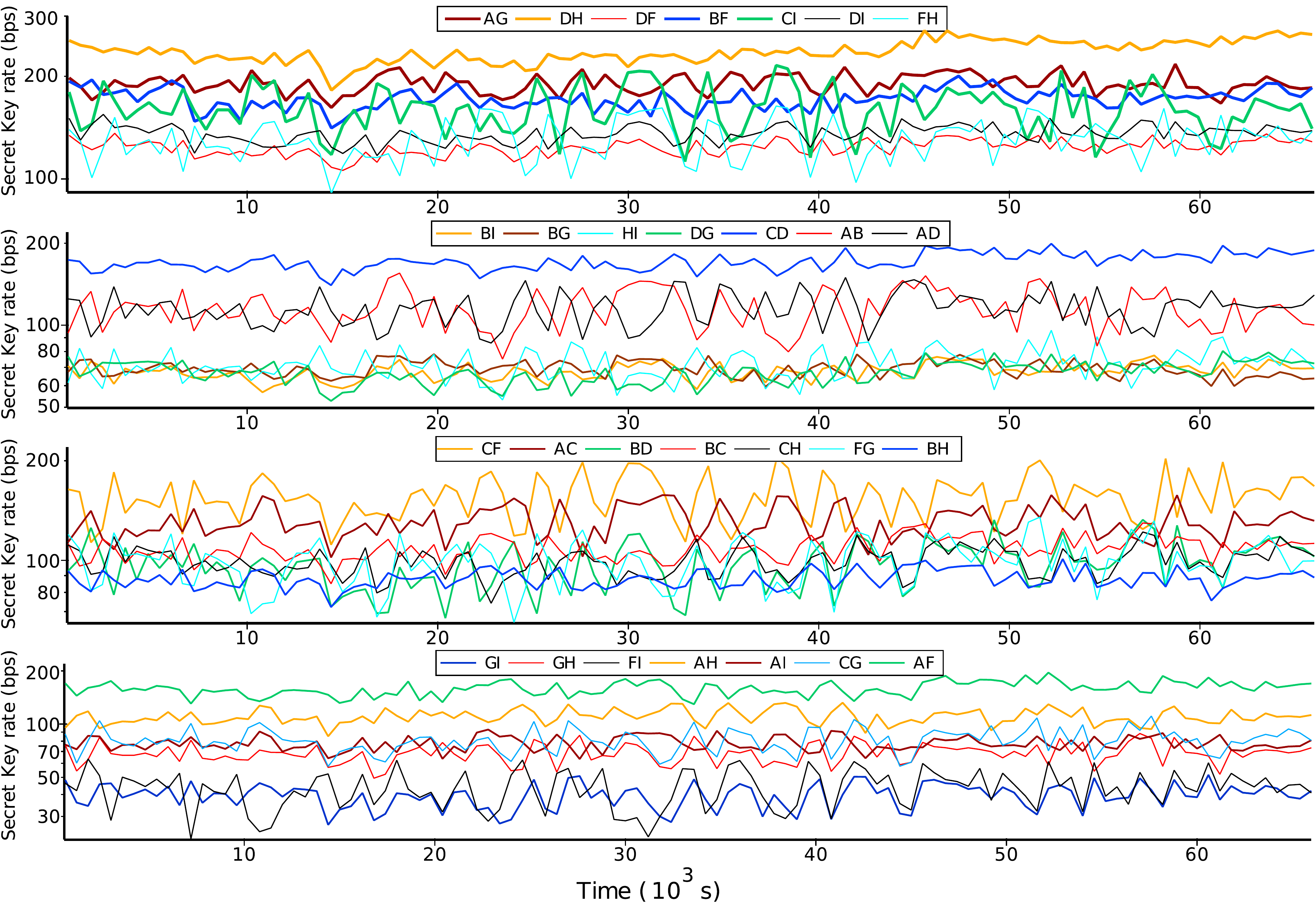}
    \caption{\textbf{Temporal cross correlation histograms between Gopi and all other users}: Each pair of users identify photon pairs  by their arrival time using $g^{(2)}$ histograms. The data shown here was collected over one hour for user Gopi (G) during the laboratory demonstration of the network. Users G and A share two sets of correlated wavelengths to enable higher key rates therefore they share two sets of $g^{(2)}$ peaks (GA$_1$ and GA$_2$). Information as to which detector(s) clicked was obscured by all users. Fig.~\ref{fig:Histgb} shows the histograms between each pair of detectors for the users G and B however, this more detailed graph contains information about the measurement outcome and cannot be used to generate a key.
    }
    \label{fig:g2}
\end{figure*}

Using a low-cost design for the polarization analysis module at each node, any pair of communicating users (say Alice and Bob) obtain three peaks in their temporal cross-correlation histogram $g^{(2)}$ between each detector of A and B for each correlated pair of wavelengths they share (see Fig.~\ref{fig:g2} and Fig.~\ref{fig:Histgb}).  {Under normal operation in the absence of an eavesdropper, t}he central peak corresponds to all measurements where A and B chose the same measurement basis, while the side peaks correspond to A and B choosing different measurement bases. {Since we do not explicitly note down the basis choice, we must assume that the BS ensures that we measure in both necessary measurement bases. In general it is sufficient to assume that the BS has a bound on its splitting ratio. (See supplementary material for more details on the security)}. 
% {The foot note does not display correctly}
%\com{@Seastian in the mathematica script you had to simulate the heights of the three peaks, can you incorporate a non 50:50 BS? Perhaps that gives us grossly unequal left and right peaks as well? Because a 90:10 splitting ratio (say) would bias the visibility measurement to the HV (say) basis.}  
%\com{I would put the next sentence to the methods section, since it's  a technicality .}
We note that each user only shares the time of a detection event and not which detector clicked as required by the protocol. Since the two detectors used by a user can have different delays or jitters, every user must characterize their setup and modify the time tags before they are shared.

\section*{Discussion}
We have successfully realized a complete entanglement based quantum communication network with improved scaling, traffic management, and long distance links via deployed fiber throughout the city. We have shown the effectiveness of a new and improved network architecture. Our fully connected scheme can be modified at the software or hardware level to create any desired sub-graph. Further, by multiplexing states intended for several users into a single fiber and demultiplexing them later on, our architecture can easily support any desired complex network (see supplementary material).

{As the number of users increases, the QNSP can choose to use additional wavelength channels, which (up to a limit based on noise counts as discussed in the supplementary material) minimally impacts the key rates of all existing users. This detrimental impact can be completely negated by users selectively detecting only the desired wavelengths. %Further, this would convert the network into a user controlled access network since the users . 
Alternatively, to increase the number of users, the QNSP can use  additional beamsplitters which would reduce key rates but drastically increase the number of users on the network with the fewest additional wavelength channels. However, this would irrecoverably affect the key rates of all users. In our network, the physical topology grows linearly with each additional user requiring only one additional PAM and fiber. Photons intended for multiple users can also be multiplexed through the same fiber to optimist cost/convenience of distribution. 
We note that the network is also capable of producing all possible sub-graphs, adding or removing users and changing the allocation of premium connections without altering the source of entanglement.}

Detectors are a significant resource for individual users
 therefore, we significantly lower the financial cost per user by sending several channels onto the same detector at the expense of a slight increase in QBER. However, this can be mitigated by demultiplexing the signal on each detector to multiple detectors. 
The all passive trusted-node-free implementation could allow us to use active switching to incorporate additional functionality such as %interconnection of networks, 
traffic and bandwidth management, Software Defined Networking etc.
The {$\sim$17\,km, or more,}  range of the network, as we have demonstrated, is more than enough to create Local and Metropolitan Area Networks interconnecting end users throughout a city or building. This range can be further extended by repeaters, reduced detector hitter, wavelength demultiplexing to several detectors and wavelength selective switching to the same detector.

\begin{figure}
    \includegraphics[width=0.95\columnwidth]{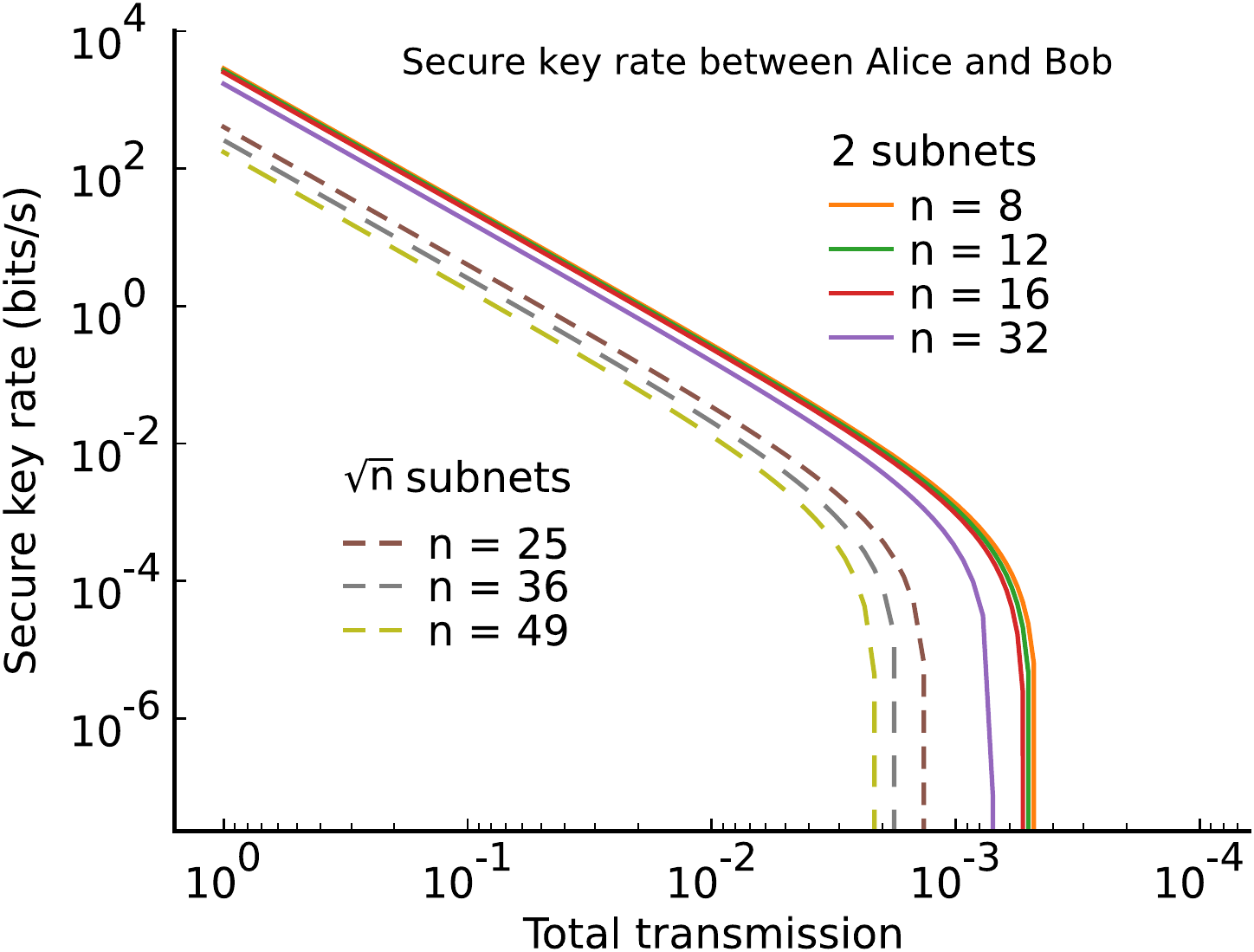}
    \caption{\textbf{Simulation showing the scalability of the current network topology}. The graph shows the secure key rate between two users who do not share a premium link. The simulation used detectors with 100\,ps jitter and 70\,\% detection efficiency. Further, the source was assumed to produce $10^5$ pairs per second per correlated wavelength pair with a heralding efficiency of 20\,\%. 
    The \textbf{solid lines} represent topologies with $n$ users split between 2 subnets using only 2-fold beamsplitters. \textbf{The dashed lines} represent the use of $\sqrt{n}$-fold beamsplitters to create $\sqrt{n}$ subnets. 128 wavelength channels are needed for the $n = 32$ 2-subnet topology, while only 84 wavelength channels are needed for the $n = 49$ 7-subnet topology. The reduced key rates of the $\sqrt{n}$-subnet topologies, despite having fewer wavelength channels, is due to the loss introduced by the $\sqrt{n}$-fold beamsplitters.
    \label{fig:numnodes}}
\end{figure}

The number of users that can be connected to a given network is limited by available resources, loss, and the marginal increase in error rate with each simultaneous connection established (with a given detector). The error rate is theoretically limited by the probability that uncorrelated photon detection events can accidentally occur in any given coincidence time window. In terms of resource, the network scales linearly with respect to user hardware and {number of deployed fibers.} On the service provider's side, the number of wavelength channels can be increased drastically by using closer-spaced narrower-band  WDMs and by generating broader band downconversion. For example, periodically poled fibers are a very promising method of the latter~\cite{Chen:18}. Using polarization preserving methods of multiplexing, such as an on chip design, would eliminate the need for most Fiber Polarization Controllers (FPCs).
We were able to demonstrate this by connecting several extra kilometers of fiber to many users and maximizing the key rate using only the FPCs on each user's module instead of the FPCs of the service provider. Further, simulations show that the network topology can be extended to 32 (49) users divided into 2 (7) subnets while maintaining a reasonable secure key rate (see Fig.~\ref{fig:numnodes}).

The long term goal of a full fledged quantum internet requires quantum communication networks that support other forms of quantum information processing or other quantum technologies. The vast majority of such applications rely on the distribution of entanglement between several nodes making the current architecture an ideal candidate for further study.

\section*{Materials and Methods}\label{sec:methods} 
\paragraph*{Quantum Network Service Provider:}
The network consists of the Quantum Network Service Provider (QNSP), distribution fibers and users. 
The QNSP is comprised of a photon pair source set up to prepare bipartite polarization entangled states and a Multiplexing Unit (MU). 

The source is pumped by diagonally polarized light $|D\rangle$, from a CW pump laser emitting at \unit[775.1085]{nm}  which passes through a dichroic mirror and a polarizing beamsplitter (PBS) that defines the input and output of a Sagnac loop (see Fig.~\ref{fig:source+wdms})~\cite{Kim2005,lim2008,fedrizzi2007wavelength}.
The horizontally (vertically) polarized pump light thus propagates anti-clockwise (clockwise) inside the loop. A half-wave plate (HWP) after the transmission port of the PBS rotates the polarization by $90\degree$ to vertical. This allows for injecting light in both directions of a {5}\,cm-long Magnesium Oxide doped periodically poled Lithium Niobate (MgO:ppLN) bulk crystal with a poling period of 19.2\,$\mu$m in which vertically polarized pump photons are converted to vertically polarized signal and idler photon pairs through type-0 spontaneous parametric down-conversion, i.e. $|V\rangle \rightarrow |V_{\rm{\textit{s}}}\rangle|V_{\rm{\textit{i}}}\rangle$. The photon pair contribution in the clockwise direction is rotated by the HWP and therefore becomes $|H_s\rangle|H_i\rangle$. Both contributions are then coherently combined at the PBS and  {are} isolated from the pump light  {by} the dichroic mirror.  {Ideally} creating maximally polarization-entangled state {s}  {for each set of} two different wavelengths, $\lambda_1$ and $\lambda_2$, located symmetrically about the central wavelength at {\unit[1550.217]{nm}}: 
\begin{equation}\label{eq:state}
|\Phi\rangle =\frac{1}{\sqrt{2}}( |V_{\lambda_1}V_{\lambda_2}\rangle + |H_{\lambda_1}H_{\lambda_2}\rangle)\,.
\end{equation}

The MU consists of off-the-shelf dense wavelength division multiplexer filters (DWDM). In addition, the MU also has a set of 50:50 fiber beamsplitters. These components were spliced together to distribute photon pairs from the source to each of the 8 distribution fibers as shown in Figs~\ref{fig:layers} and~\ref{fig:source+wdms}.

%\subsection{Network preparation}

The spatial mode containing the signal and idler photons from the source was coupled into one single mode fiber and spectrally split by a thin-film DWDM  {from Opneti} (with a channel spacing  and nominal full width of  \unit[100]{GHz}) into 32 channels as defined by the International Telecommunication Union (ITU) in G.694.1. Our QNSP consists of one 32 channel DWDM (of which only 16 are used) exhibiting insertion losses per channel between 0.6 -- 2.5\,dB (and a Polarization Dependent Loss (PDL) $<$
{\unit[2.5]{dB}} according to the data-sheet), 16 add/drop DWDMs with $\sim${\unit[0.5]{dB}} insertion loss (PDL $<$ {\unit[0.1]{dB}}) and 8 standard fused couplers with insertion loss below {\unit[3.4]{dB}} (PDL $<$ {\unit[0.1]{dB}}). { Optical multiplexers form the foundation of DWDM networks deployed by the telecommunication industry. There is currently two main technologies used in the industry, thin-film filters (TFF) and arrayed waveguide gratings (AWG). TFFs function by filtering wavelengths serially and are designed to transmit a specific wavelength while reflecting all others. They are made of a concatenation of interference filters each fabricated with a different set of dielectric coating. AWGs are single-stage filters that deploy planar waveguide technology consisting of free propagating regions interconnected by waveguides. The waveguides have different length leading to constructive or destructive interferences in the output and thus multiplexing/demultiplexing. Due to their low-cost and robustness against thermal fluctuations, we have chosen TFF to build our QSNP. Also, the main advantage of AWGs over TFF is that the parallel multiplexing approach is more conductive to high channel-count applications, which is not relevant for quantum signal levels.}

We selected 16 {wavelength} channels symmetrically with respect to the degeneracy wavelength of {\unit[1550.217]{nm}}, which corresponds to ITU channel 34 (see Fig.~\ref{fig:source+wdms} inset a) ). On the red side of the spectrum we used ITU frequency channels 26-33 and channels 35-42 on the blue side. Due to the well defined pump wavelength of the CW laser and energy conservation during down-conversion, we obtained polarization entanglement between pairs of channels (26 \& 42, 27 \& 41, 28 \& 40, and so on).

Each of the 16 wavelength channels is then split by a beamsplitter, such that in total 32 possible pairs of correlated photons are available. Using further add-drop multiplexers before and after the beamsplitters, four channels were combined into each single-mode fiber to every user. Since the partner photons for each channel can be found in two other fibers, each of the users now holds eight polarization-entangled connections to other users. This means, each user is connected to all the other users, featuring one doubled connection. 

Fiber Polarization Controllers (FPCs) were used to ensure that the reference frame of polarization in the source is (nearly) identical to that of the polarization analysis module (PAM). 
%As discussed in the main text, in order for 8 users to share at least one correlated wavelength channel with every other user, we provide 4 specific channels to each user. Using beamsplitters each wavelength channel or group of wavelength channels were split to create the two sub-nets. 
%This is accomplished by doubling 4 channels with a BS and then cleverly multiplexing them back with the other 12 channels after they have been multiplexed 3 at a time and then doubled. 
%Due to the use of the beamsplitters after combining several channels into one fiber, 
It was not necessary to compensate all channels independently. Similar wavelengths were compensated together.  At the end each user received 4 channels (see Figs~\ref{fig:layers} and~\ref{fig:source+wdms}) via a single distribution fiber and used a polarization analysis module to measure in the HV or DA polarization basis. 

The distribution fibers were all single mode for 1550\,nm but of varying lengths and specifications. Several of the fibers were deployed across the university and the city of Bristol. We conducted two experimental runs, the first with short distribution fibers and the second with varying link lengths as shown in Table~\ref{tab:length}.

\begin{table}[htb]
%\centering

    \begin{tabular}[b]{c|c|c|c|c|c|c|c|c}
      & Bob   & Chloe & Dave  & Feng  & Gopi  & Heidi & Ivan  \\\hline
Alice        & 12642 & 13095 & 16971 & 14257 & 12642 & 14256 & 15735 \\
Bob          &       & 473   & 4350  & 1636  & 20    & 1634  & 3113  \\
Chloe        &       &       & 4803  & 2089  & 473   & 2087  & 3566  \\
Dave         &       &       &       & 5965  & 4350  & 5963  & 7442  \\
Feng         &       &       &       &       & 1636  & 3250  & 4728  \\
Gopi         &       &       &       &       &       & 1635  & 3113  \\
Heidi        &       &       &       &       &       &       & 4727  \\
%Ivan  &       &       &       &       &       &       &       &       \\
\hline
\end{tabular}

\raggedright
\caption{\textbf{Length of each link in meters} in the metropolitan network shown in meters. Bob and Gopi were users separated by 10\,m of fiber each from the QNSP. Alice and Dave were connected to $\sim$12.6\,km and $\sim$4.3\,km spools of fiber. The remaining users were connected via loop-back to various locations across the city of Bristol as shown in Fig.~\ref{fig:source+wdms}. Each link was characterized by an OTDR and the measurements shown are the link distances in fiber between each pair of users. }

    \label{tab:length}
\end{table}

\paragraph*{Users:}
Each user in the network is equipped with the polarization analysis module (PAM) and two single photon detectors. The PAM enables passive switching between photon polarization measurements in two orthonormal bases (either HV or DA). A beamsplitter (BS) at the PAM's input randomly directs incoming photons either through the short optical path to the PBS and measurement in HV basis, or through the long optical path with an achromatic half-wave plate providing a $45\degree$ polarization rotation and the same PBS for measurement in DA basis. The difference between the long and short free-space paths corresponds to 3.7\,ns of time delay between, resulting in polarization analysis in different time bins\cite{Zhu:19}.

We designed the PAM to be completely passive, compact, portable, simple and cheap to mass-produce and align, but still robust and stable. 
%\com{I don't see why advertising software packages is important and vote to delete the next  sentence.} For the optical design of fiber collimators/couplers we used the Physical Optics Propagation (POP) tool in ZEMAX software and for the mechanical design of the PAM setup with all kinematic mounts and base plates we used MegaCAD software.
%
At each PAM's input and outputs to two detectors, SMF28e single mode fibers are connected to collimators/couplers with custom produced 15.7 mm effective focal length (at 1550~nm) SF11 glass singlet lenses (AR coated for 1500-1600~nm), with x-y, tip/tilt and focus adjustment. Cube BS and PBS are mounted on kinematic platforms for rotation and tip/tilt adjustment. The achromatic HWP is mounted in a manual precision rotation mount. The long optical path was realized using unprotected gold mirrors on tip/tilt kinematic mounts. 

Commercially available BS, PBS, HWP and unprotected gold mirrors were used (Thorlabs BS012, PBS104, AHWP05M-1600 and PFSQ10-03-M03).
BS characteristics are $T=(56 \pm 8)\%$, $R = (44 \pm 8)\%$ depending on input polarization and orientation angle. 
The PBS in use have extinction ratios $T_{p}:T_{s} > 1000:1$, $R_{s}:R_{p}$ roughly between 20:1 to 100:1, transmission efficiency $T_{p} > 90\%$ and reflection efficiency $R_{s} > 99.5\%$ {, where $T$ and $R$ represent Transmitted and Reflected ports respectively and the subscripts $s$ and $p$ represent the $s$ and $p$ polarized components}. Achromatic HWP retardance accuracy is $<\lambda/300$.
Complete production of lenses and optomechanics as well as assembly was done at the Ru\dj{}er Bo\v{s}kovi\'{c} Institute, in the optical and mechanical workshops of the Division of Physical Chemistry.

The PAM outputs are fiber coupled and launched into 2 single photon detectors. We used 15 superconducting nanowire single--photon detectors (SNSPDs) from Photon Spot with  detection efficiencies ranging from $\sim$70 to 90\%, a jitter of  between $\sim$80 to 60\,ps (including the measurement device) and dark counts of $\sim$1 kHz and one InGaAs avalanche single photon avalanche detector (SPAD) from ID Quantique, model ID230, which has 20\% efficiency, $\sim$100 ps jitter and dark counts of $\sim$0.05 kHz.   After detection the optical signal is converted into an electronic pulse and read out at a 18-channel time tagging module (Swabian instrument model Time Tagger Ultra). Using a laptop we performed an  on-the-fly computation of coincidences, basis reconciliation and secure key rate estimation for all 28 QKD links.

\paragraph*{Secure key generation:}
Due to the design of PAMs and the Continuous Wave (CW) pump of the source, we only know whether  the measurement basis choice (of a pair of users) was matched or not. Here the information of which detector clicked directly encodes the measurement outcome without revealing the measurement basis choice. Thus,  {we cannot know the basis in which a detected photon was measured}. Suppose the time delay between Alice's HA detector to Bob's HA detector is longer than the delay between Alice's VD detector and Bob's VD detector, then by looking at the $g^{(2)}$ histogram, Eve can identify two different delays, which means that Eve can guess what the measurement outcome was and thus what the key bit could be. Thus, all users must merge the time tags of all detectors into a single data set without which-channel or which-detector information. %They then exchange this data} to synchronize the time offset between different correlated detector pairs. %This process also removes the channel information.

Then, all users exchange their merged time-tags with each other via the authenticated public channel(s).  After the time-tags are shared among all users, they calculate a temporal cross-correlation histogram ($g^{(2)}$)  to find the coincidences. 
%As the channel information is kept private, u
 {U}sers assign ``0'' or ``1'' to the measurement outcomes where they both detect a photon within the coincidence window and happened to measure in the same basis.  
%Eve could learn the position of matched pulses, but doesn't know the value of the matched key bit due to lack of channel information. 
After obtaining the sifted keys, all users perform the error correction and privacy amplification procedures to extract the final secure key. 
 {Assuming that each pair of users has been able to identify and include in their sifted key only those rounds in which they happened to measure in the same basis (see security considerations in the supplementary material for more details), their} final secure key length $n_f$ can be calculated by
\begin{equation}
\label{equ:finalkeylength}
    n_f \geq n_s\left[1-H_2\left(e_p^U\right) -  {f} H_2\left(e_b\right)\right]\,,
\end{equation}
where $n_s$ is the sifted key length, $e_b$ is the measured quantum bit error rate, $e_p^U$ is the estimated upper-bound of phase error rate, $H_2\left(x\right)$ is the binary Shannon entropy and  {$f$ is the error correction inefficiency (assumed to be 1)}.

Since we could not divide our sifted key into two individual bases ($Z$ and $X$), here we analyze the phase error rate in the mixed basis case. Given failure probability $\xi_{ph}$, the upper-bound of phase error rate can be estimated by
\begin{equation}
\label{equ:phaseErrRate}
    e_p^U= \alpha e_b +  \left(1+\alpha\right)\sqrt{\frac{\ln 4 - {2}\ln {\xi_{ph}}}{n_s}}\,,
\end{equation}
where $\alpha \geq 1$, and the phase error probability is $\alpha$ times bit error probability. In our experiment, we used passive measurement modules with 50:50 beamsplitters to perform unbiased measurement basis choices, which results in $\alpha=1$~\cite{Fung2010}. 
Similar arguments can be made to show security even in the case of bias in the measurement basis choice. From the experimental data, we can infer this bias under the assumption that the fiber-coupling for both detectors of the PAM is equal. 
%\com{... or is there a way to deal with a different basis choice bias for each detector?} 
Assume that the basis choice bias of Alice and Bob is $p_Z^A$ and $p_Z^B$, then the coincidence counts of the left histogram peak between a pair of users (as seen in the histograms of Fig.~\ref{fig:Histgb})  is related to $p_Z^A(1-p_Z^B)$, the coincidence counts of the middle peak are related to $p_Z^A p_Z^B + (1-p_Z^A)(1-p_Z^B)$, the coincidence count of the right peak is related to $(1-p_Z^A)p_Z^B$. Therefore, when the basis choice is biased, one could first measure $p_Z^A$ and $p_Z^B$  by monitoring the coincidence counts in  the left, middle and right peaks. Then, one can estimate $\alpha$ using these values of  $p_Z^A$ and $p_Z^B$.  {More details on the security of this scheme can be found in the Supplementary material.}

% Your references go at the end of the main text, and before the
% figures.  For this document we've used BibTeX, the .bib file
% scibib.bib, and the .bst file Science.bst.  The package scicite.sty
% was included to format the reference numbers according to *Science*
% style.

%BibTeX users: After compilation, comment out the following two lines and paste in
% the generated .bbl file. 
%\bibliographystyle{ieeetr}
%\bibliography{Qnet}

\section*{Acknowledgments}
The research leading to this work has received funding from the Engineering and Physical Science Research Council (EPSRC) Quantum Communications Hubs EP/M013472/1 \& EP/T001011/1 and equipment procured by the QuPIC project EP/N015126/1. We acknowledge the Ministry of Science and Education (MSE) of Croatia, contract No. KK.01.1.1.01.0001.
We acknowledge financial support from the Austrian Research Promotion Agency (FFG) project ASAP12-85 and project SatNetQ 854022.
This work was partially supported by the European Union’s Horizon 2020 research and innovation programme under the Marie Sklodowska-Curie grant agreement number 675662 (QCALL).
 {AQ was funded through the Quantum Engineering Centre for Doctoral Training EP/LO15730/1.}

We gratefully acknowledge assistance from Ashley Bligh and Ian Stubbs of the University of Bristol networks team as well as Graham Marshall of QET labs, University of Bristol for provisioning parts of the fiber network used in this research.
We thank Vitomir Stani\v{s}i\'{c} of RBI-ZFK Workshop for his help designing and making the mechanical components for the user PAMs. We also thank Roland Blach of IQOQI for help making the FPCs and Abdul Waris Ziarkash for helpful discussions.
All data from this publication is stored for at least 10 years on the University of Bristol's Research Data Storage Facility (RDSF). The processed data for the findings in this paper is available publicly from the RDSF, the raw data consisting of timetag files is too large to host publicly and available from the corresponding author on request.

The Source was built by SN and SW, the nodes by ML, ZS, SKJ and SW and the remaining setup by DA and SKJ. The network topology/architecture was conceived by SKJ with help from SW and DA.  {MR and GCL contributed the security proofs.} Security analysis of the data was performed by BL and TS. {Test, monitoring and secure key analysis} software was written by BL. The Project was supervised by RU, JR and MS who coordinated the 3 teams while the whole project was lead by SKJ. Simulations were performed by AQ and SKJ with help from SN and SW. LK provided logistic support and helped with the detection of single photons. ML, and ZS produced the low-cost optics and optomechanics. The experiment was performed by SKJ, DA, SW, SN, ML, LK, ZS, TS, LB. The paper was written by  SKJ with the help of DA, SW and BL. All authors discussed the results and commented on the manuscript. 

Competing Interests: The authors declare that they have no competing interests.

\section*{Supplementary Material}\label{appendix}

\paragraph*{Scalability:}\label{scale}
%\linenumbers

We have demonstrated an eight-user quantum communication network enabled by a single source and wavelength/BS multiplexing. The number of wavelength channels available depends on the available WDM technology and the bandwidth of the polarization entangled photon pair source. The current source has a bandwidth of $\sim$60\,nm ($\sim$7492\,GHz) and the WDM channels used have a $\sim$0.8\,nm (100\,GHz) bandwidth. This limits us to a maximum of $\sim$75 wavelength channels. However, broader bandwidth sources (such as supercontinuum  based sources of polarization entanglement) and or closer spaced WDM channels (such as the upcoming 10 GHz DWDM standard) would allow at least a few hundred channels.

Thus the real limit to the maximum number of users in such a quantum network is the QBER, specifically the contribution from ``accidentals''. Given a particular photon flux incident on the detectors of two users, there is a probability that two uncorrelated detection events happen to occur within the chosen coincidence window. Such coincidences are called accidentals. In our experiment, a user opens a different coincidence window for every incident wavelength channel as needed because of the  uncompensated propagation times of different wavelengths via the fibers and DWDM channels.
Thus increasing the number of wavelengths a single user receives increases the accidental rate and the QBER, effectively reducing the secure key rate.
The QNSP can correct for the propagation delays between different wavelength and BS, when all users are connected by fixed lengths of fiber, dramatically increasing the secure key rates because each user will then need to consider only a single coincidence window. Similarly the users have two strategies to increase their secure key rates: use multiple detectors for each measurement outcome, each of which detects fewer wavelength channels; or selectively choose which wavelengths are incident on their detectors based on the connection(s) desired with select other users.
In the absence of such methods, minimizing the number of multiplexed channels is the best way to increase secure key rates. 

By using $k$-fold beamsplitters, we can create $\frac{n}{k}$ subnets between $n$ users. Each subnet forms a fully connected network with WDMs only using $\frac{n}{k}\left(\frac{n}{k} -1 \right)$ wavelength channels. Additional wavelength channels are needed to interconnect the different subgroups. This can be done in two ways: First, we treat each subnet effectively as a single user in a $k$ user network and create a fully connected network with $k(k-1)$ additional wavelength channels. This implies using  $\frac{n}{k}$-fold beamsplitters to distribute each of these extra $k(k-1)$ channels to all $\frac{n}{k}$ users in a subnet. When considering the link between just two users, beamsplitters can be viewed as losses. Thus mixing $k$-fold and $\frac{n}{k}$-fold beamsplitters can result in significantly different coincidence rates between different sets of channels/users. Consequently, the optimal pump power (i.e. pair rate emitted by the source) will be different for various channels. With a single source, we cannot optimist this independently which in turn leads to sub optimal key rates. Thus in the second method, we can impose the constraint that all wavelength channels are split using only $k$-fold beamsplitters. Here, each subnet requires $\frac{n}{k^2}$ wavelengths and the network requires a total of $\frac{n}{k}\left(\frac{n}{k} -1 \right) + \frac{n}{k^2}k(k-1)$ wavelength channels. We note that the above formulas are valid only when $n$, $k$, $\frac{n}{k}$ and $\frac{n}{k^2}$ are all integers. To create networks with any integer number of users, it is possible to create a larger network that satisfies the above conditions and not connect all users as and how required.

With a fixed pump power of the entangled photon source, the net effect of BS multiplexing a correlated wavelength pair is akin to ``time sharing'' the key. When considering just two users, it can be thought of as additional loss. On the other hand, wavelength multiplexing introduces a new source of key.  Both techniques adversely affect the signal to noise ratio however, the additional noise due to  more wavelength channels can be avoided by filtering.  Thus a good topology makes use of a significant amount of wavelength multiplexing supplemented by BS multiplexing.
Fig.~\ref{fig:numnodes} shows the secure key rate for different numbers of users as a function of the transmission loss. For convenience, we have focused on the use of 2-fold beamsplitters (solid lines) and networks where $n = k^2$, using $k$-fold beamsplitters and $k$ subnets (dashed lines).
Maintaining a reasonable secure key rate, within the constraints of the current experiment, is possible for both 32 users in 2 subnets and even 49 users in 7 subnets. This demonstrates how our network architecture can be used for very large and complex Local Area Quantum Networks.

%\nolinenumbers

\paragraph*{Adapting the network for practical use-cases:}\label{use}
%\linenumbers
In our proof of principle demonstration, the source of polarization entangled photon pairs was adjacent to the Multiplexing Unit (MU) that distributed these pairs among the users. The users were in turn each connected to the MU by single long fibers. In a real world scenario, the MU need not be adjacent to the source nor must the MU be a single unit. Photons from the source can be split and combined in any physical location or in several locations to take advantage of available fiber infrastructure. In our experiment the source was separated from the MU by a 8\,m long fiber and each component in the MU was separated by between 2 and 7.5\,m of optical fiber.

Further, our network topology allows the flexibility needed for a wide variety of use-cases. Longer distance, or higher speed links can co-exist on the same network; these can be implemented by using multiple detectors on certain nodes where each detector measures some of the incident wavelengths. Access networks are often preferred for low bandwidth use-cases, however users sacrifice anonymity when they request the Quantum Network Service Provider (QNSP) to establish a connection with another chosen user. Here, our topology is capable of supporting an anonymous access network controlled by each user instead of the QNSP. 

%\nolinenumbers

\paragraph*{Optimizing the key rate:}\label{key}

It is possible to significantly improve the key rate beyond what is shown in the main paper in several ways. First, increasing the pump power increases the number of photon pairs. Given the detector jitter, losses and QBER there exists an optimum pump power at which the 
secure key rate (in bits per second) is maximized. Fig.~\ref{fig:pow} shows the key rate of all 28 links in the network measured at 9 different pump powers. Note that when the photon flux is excessive, a secure key cannot always be generated. This is because of the increased contribution of uncorrelated singles to the QBER via accidental counts. Reducing the detector jitter is thus the best way to further increase the key rates.
Using a single source for the entire network limits our control over the individual pair production rates for each correlated wavelength pair. Thus using different types of detectors strongly influences the optimal pump power. In addition to different detectors, the alignment of individual PAMs of each user and the FPCs contribute to the overall network performance.  
Second, using a pulsed pump as discussed in the supplementary material of Ref~\cite{Soeren2018} would help reduce the QBER and significantly increase the key rates.
Third, in our experiment, we utilized several manual Fiber Polarization Controllers (FPCs) which were needed to maintain the polarization entangled state at each stage of the multiplexing, demultiplexing, beamsplitting and distribution. For expediency and to demonstrate the success of our network topology, we considered it sufficient when each of the FPCs were aligned with $>$ 97\,\% visibility. A better fiber neutralization would have resulted in improved key rates.
Lastly, in an attempt to keep the costs of each user's Polarization Analysis Module (PAM) to a minimum, we used readily available sub-par components which we estimate contribute to the overall QBER by up to an additional 1\,\%.

\paragraph*{ {Security considerations:}}\label{sec:security}
In this section, we provide a more detailed analysis of the security of the implemented protocol. The protocol used in our experiment is slightly different from the original BBM92 in the following aspects. First, while the random choice of measurement basis is performed passively in our setup by a 50:50 beamsplitter, the users cannot tell in which basis they have done the measurement. Instead, if after correcting for their time offsets, they both detect a photon within a coincidence window of width $\tau_c$, they assume that they have both used the same basis. In this mixed-basis case \cite{Fung2010}, the estimation of the phase error rate from the observed bit error rate must be done with caution. In particular, because, in our setup, the employed beamsplitters in the receiver units may not be exactly 50:50, one should account for its effect on the secret key rate.

Here, we use a simplified picture of the protocol to account for the above two issues in our security analysis. We only consider two nominal users, Alice and Bob; the same argument holds for any pair of users in our setup. Without loss of generality, we assume that  Alice and Bob share the same time reference and that the transmission delay between the source and each of the two users is zero. Now consider a particular pattern of detection events that corresponds to a certain sifted key bit. That is, suppose Alice and Bob have detection events, respectively, at time $t_{A}$ and $t_{B}$ such that $|t_{A}-t_{B}|<\tau_c/2$. We refer to such an event as a coincidence with time offset at the receiver, $\Delta_{\rm Rx}$, equal to zero. (More generally, we have a coincidence event at a nonzero $\Delta_{\rm Rx}$ if $|t_{A}-t_{B}-\Delta_{\rm Rx}|<\tau_c/2$.) There should then be a transmitted signal to Alice (Bob) at time $\tau_{A(B)} \in \{t_{A(B)},t_{A(B)}-\Delta\}$, where $\Delta$ is the time difference between the long and short optical paths, used for X and Z basis measurements, in the PAMs. For simplicity, we assume the time delay in the short path is zero.

The key point in our security proof is that, so long as $\Delta \gg \tau_c$, the only detection events that can be used for secure key extraction are those for which $|\tau_A - \tau_B|< \tau_c/2$. We refer to such an event as a coincidence with time offset at the transmitter, $\Delta_{\rm Tx}$, equal to zero. More generally, we have a coincidence event at a nonzero $\Delta_{\rm Tx}$ if $|\tau_{A}-\tau_{B}-\Delta_{\rm Tx}|<\tau_c/2$. As we explain below, the detection events that originate from transmitted signals with $\Delta_{\rm Tx} = \pm \Delta$ can easily be manipulated by a potential eavesdropper to give us insecure detection events. Note that in the trust-free QKD setting that we are considering, we cannot assume that the source is trustworthy. Even if we make this assumption, an eavesdropper can block the trusted source output, and, instead, send her own signals to the users. Now, imagine such an eavesdropper is sending an A-polarized photon to Alice at time $t_A-\Delta$ and an H-polarized photon to Bob at time $t_B$, with $|t_{A}-t_{B}|<\tau_c/2$. Then a detection event at times $t_A$ and $t_B$ would correspond to the same bit but different bases, while Alice and Bob would falsely assume that these are in the same basis as expected from an honest source. However, the eavesdropper can tell without any error the bits assigned to the sifted key in such a case. In other words, in the terminology of the GLLP analysis \cite{gottesman2004security}, these bits are {\em tagged}.

Luckily, such an eavesdropping attempt would leave a footprint, which could be used to estimate the amount of information that has leaked to Eve. In the example above, a signal generated at $t_A-\Delta$ may also take the shorter path and cause a click at the same time on Alice's side, while the signal generated at $t_B$ takes the longer path and causes a click at $t_B+\Delta$. Having a coincidence event with a time offset $\Delta_{\rm Rx} = 2\Delta$ would not have been expected if the signals sent to Alice and Bob are generated at the same time. We will use the collected data on the latter events to bound the number of tagged sifted key bits.

Based on the above, for any coincident event with $\Delta_{\rm Rx} = 0$, there are only three possible transmission time offsets, namely, $\Delta_{\rm Tx} \in \{-\Delta,0,\Delta\}$. The sifted key bits for which Eve chooses $\Delta_{\rm Tx} = \pm \Delta$ are tagged: We should assume that Eve can fully learn them without introducing any errors. The sifted key bits for which Eve chooses $\Delta_{\rm Tx} = 0$ are untagged, and they can be regarded as having arisen from an execution of a standard BBM92 in which Alice and Bob have been able to postselect the detected rounds in which they have used the same basis, but not to learn their specific choice of basis for each round. We assume that the signals received in these rounds are in a qubit space (i.e., a polarized single photon); if they are not, one can still prove security by using the techniques of \cite{tsurumaru2008security,beaudry2008squashing} and assigning a random sifted bit to events in which more than one detector clicks in a particular round. To estimate the amount of secret key that can be extracted, we then need to obtain: \textbf{(1)} a lower bound on the number of untagged bits $N_{0,0}$ in the sifted key, where $N_{R,T}$ denotes the number of events in which $\Delta_{\rm Rx}=R$ and $\Delta_{\rm Tx}=T$; and \textbf{(2)} an upper bound on the phase error rate of these bits, which we denote as $e_p$.
		
\noindent\textbf{(1) Lower bound on $N_{0,0}$:} For bounding $N_{0,0}$, we use the fact that when Eve chooses $\Delta_{\rm Tx} = \pm \Delta$, the probability of having $\Delta_{\rm Rx} = 0$  is the same as that of $\Delta_{\rm Rx} = \pm 2 \Delta$. We denote the number of coincidence events with receiver time offsets of either zero or $\pm 2 \Delta$ by $N$.  For event $n$ out of these $N$ events, we then have
	\begin{equation}
	\label{eq:pr-relationship}
		\Pr [\Delta_{\rm Tx}^{(n)} \in \{\Delta, -\Delta \}, \Delta_{\rm Rx}^{(n)} = 0] = \Pr [\Delta_{\rm Tx}^{(n)} \in \{\Delta, -\Delta \},  \Delta_{\rm Rx}^{(n)}  \in \{2\Delta, -2\Delta \}] \leq \Pr [\Delta_{\rm Rx}^{(n)} \in \{2\Delta, -2\Delta \}],
	\end{equation}
	where the superscript $(n)$ specifies the value of the time offset parameters for the $n$th event. By Azuma's inequality, we have that
	\begin{equation}
	\label{eq:azumapr}
	\begin{gathered}
		\sum_{n=1}^N \Pr [\Delta_{\rm Tx}^{(n)} \in \{\Delta, -\Delta \}, \Delta_{\rm Rx}^{(n)} = 0] \geq  N_{0,  \Delta} + N_{0, - \Delta}  - \delta, \\
		\sum_{n=1}^N \Pr [\Delta_{\rm Rx}^{(n)} = \in \{2\Delta, -2\Delta \}] \leq  N_{2 \Delta} + N_{-2 \Delta} + \delta,
	\end{gathered}
	\end{equation}
	where each of the bounds fails with probability $\varepsilon$, $\delta = \sqrt{2 N \ln \varepsilon^{-1}}$ is the deviation term, and $N_{R}$ is the total number of detections for which $\Delta_{\rm Rx} = R$. The conditioning on the outcome of the previous detections has been omitted from all probability terms for simplicity. Combining \cref{eq:pr-relationship} and \cref{eq:azumapr}, we have that
	
	\begin{equation}
		N_{0,\Delta} + N_{0,-\Delta} \leq N_{2 \Delta} + N_{-2 \Delta} + 2 \delta,
	\end{equation} 
	and, therefore,
	\begin{equation}
	N_{0,  0} = N_{0} - N_{0, \Delta} -  N_{0, -\Delta} \geq N_{0} - N_{2\Delta} - N_{-2\Delta} - 2 \delta := N_{0, 0}^{(L)}, \
	\end{equation}
	except with probability $2 \varepsilon$.
		\hfill
	
\noindent	\textbf{(2) Upper bound on $e_p$:} To bound the phase-error rate of the untagged bits (that is, those bits for which $\Delta_{\rm Tx} = 0$ and $\Delta_{\rm Rx}  = 0$), we bound the number of phase errors $N_{\rm ph}$ that Alice and Bob would have obtained in a hypothetical scenario in which they have made exactly the opposite basis choices as in the real scenario.
Let us assume that $p_Z^A = p_Z^B = p_X^A = p_X^B = \frac{1}{2}$, where $p_K^{A(B)}$ is the probability of choosing basis $K=X,Z$ by Alice (Bob). Then, the probability that Alice and Bob both measure in the $Z$ ($X$) basis in the real (hypothetical) scenario is the same as the probability that they both measure in the $X$ ($Z$) basis in the real (hypothetical) scenario. In this situation, we have that, for a given untagged round,  the probability that Alice and Bob obtain an error is the same for both the real and hypothetical scenarios. That is, its bit error probability equals its phase error probability \cite{Fung2010}.

If the measurement basis choice is biased, the two are no longer necessarily equal. Say that Alice and Bob are $\alpha$ times as likely to jointly choose one basis than the other, e.g. $p_Z^A p_Z^B = \alpha p_X^A p_X^B$ with $\alpha \geq 1$. Then, if Eve makes the $X$-basis error probability larger than the $Z$-basis error probability, the phase-error probability will be larger than the bit-error probability. Still, one can easily show that the phase-error probability will be at most $\alpha$ times larger than the bit-error probability \cite{Fung2010}. Then, we have that
\begin{equation}
\begin{gathered}
\label{eq:err-relationship}
\Pr [\Delta_{\rm Tx}^{(n)}  = 0, \Delta_{\rm Rx}^{(n)} = 0, \text{phase error}] \leq \alpha \Pr [\Delta_{\rm Tx}^{(n)}  = 0, \Delta_{\rm Rx}^{(n)} = 0, \text{bit error}] \\ \leq \alpha \Pr [\Delta_{\rm Rx}^{(n)} = 0, \text{bit error}].
\end{gathered}
\end{equation}
	And, by Azuma's inequality, we have that
	\begin{equation}
	\label{eq:azumaerr}
	\begin{gathered}
	\sum_{n=1}^N \Pr [\Delta_{\rm Tx}^{(n)}  = 0, \Delta_{\rm Rx}^{(n)} = 0, \text{phase error}] \geq  N_{\text{ph}} - \delta, \\
	\sum_{n=1}^N \Pr [\Delta_{\rm Rx}^{(n)}  = 0, \text{bit error}] \leq  N_{\text{err}} - \delta,
	\end{gathered}
	\end{equation}
	where $N_{\text{err}}$ is the amount of bits in the sifted key that have a bit error (i.e. $e_{b} = \frac{N_{\text{err}}}{N_0}$), and $N_{\text{ph}}$ is the number of phase errors, defined above. Combining \cref{eq:err-relationship} and \cref{eq:azumaerr}, we have that
	\begin{equation}
		N_{\text{ph}} \leq \alpha  N_{\text{err}} + (1+\alpha) \delta := N_{\text{ph}}^{(U)},
	\end{equation}
	except with probability $2\varepsilon$. The phase-error rate can now be simply upper-bounded by
	\begin{equation}
	\label{eq:epU}
		e_{p}^U = \frac{N_{\text{ph}}^{(U)}}{N_{0,0}^{(L)}}.
	\end{equation}
	Finally, the length of secret key that can be distilled is given by
	\begin{equation}
		n_f \geq N_{0, 0} ^{(L)} \left[1-H_2(e_{p}^U)\right] - f(e_b) N_{0} H_2(e_{b}). 
	\end{equation}
	In the methods section, we use a simpler version of the above expression in which we assume that $\Delta _{\rm Tx} = 0$ for all rounds. In this case, all sifted-key bits are untagged, which implies that $N_{0, 0} ^{(L)} = N_{0,0} = N_0 \equiv n_s$, and $e_{p}^U$ in \cref{eq:epU} reduces to \cref{equ:phaseErrRate} of the main text, with $\xi_{ph} = 2 \varepsilon$.

\renewcommand{\figurename}{\textbf{Fig.}}
\renewcommand{\tablename}{\textbf{Table}}
\renewcommand{\thetable}{\textbf{S\arabic{table}}}  
\renewcommand{\thefigure}{\textbf{S\arabic{figure}}}
\setcounter{figure}{0}
\setcounter{figure}{3}
\setcounter{table}{0}

%\clearpage

%\clearpage

\end{document}